\documentclass{svmult}

\usepackage{}
\usepackage[dvips]{graphicx}

\begin{document}
\begin{center}
{\Huge Consistency Principle in Biological Dynamical Systems}
\end{center}

\begin{Large}
\begin{center}
Kunihiko Kaneko\footnotemark[1]\footnotemark[3] and Chikara Furusawa\footnotemark[2]\footnotemark[3]{}
\end{center}
\end{Large}
\begin{center}
\noindent
\footnotemark[1] Department of Basic Science, Univ. of
Tokyo, 3-8-1 Komaba, Tokyo 153-8902, Japan\\
\footnotemark[2] Department of Bioinformatics Engineering, 
Osaka University, 2-1 Yamadaoka, Osaka 565-0871, Japan\\
\footnotemark[2] ERATO Complex Systems Biology Project, JST, 3-8-1
Komaba, Tokyo 153-8902, Japan \\
\end{center}

\begin{abstract}
We propose a principle of consistency between different hierarchical levels of biological systems. Given a consistency between molecule replication and cell reproduction, universal statistical laws on cellular chemical abundances are derived and confirmed experimentally. They include a power law distribution of gene expressions, a lognormal distribution of cellular chemical abundances over cells, and embedding of the power law into the network connectivity distribution. Second, given a consistency between genotype and phenotype, a general relationship between phenotype fluctuations by genetic variation and isogenic phenotypic fluctuation by developmental noise is derived. Third, we discuss the chaos mechanism for 
stem cell differentiation with autonomous regulation, resulting from a consistency between cell reproduction and growth of the cell ensemble.
\end{abstract}

\section{Introduction}

Biological systems generally form a hierarchy. Ecological systems consist of a population of organisms, an organism consists of an ensemble of cells, and a cell consists of interacting biomolecules. Of course, such hierarchical structures also exist in nonliving systems. Then, is there some characteristic property underlying biological hierarchical systems?
In a hierarchical system, the description of units at a lower level and a description of how they come together may lead to an understanding of the upper level. However, this bottom-up picture may not be sufficient for a complex biological system, since each unit at a given lower level is not rigidly determined but can change in adaptation to feedback from a higher level.

As an example, consider a cell in a multicellular organism, which has internal degrees of freedom and can change its chemical composition or gene expression patterns. (This is in strong contrast with an electron functioning as a hierarchical unit in a physical system). Through interactions with other cells, the characteristics of a cell are changed through the process of cell differentiation. A cell in isolation and a cell in a community sometimes exhibit different characteristics, since the importance of cell--cell interactions is so significant. A cell, and a tissue as an ensemble of cells, mutually determine their character. In other words, the character of a unit (e.g., a cell)
is determined not independently but is changed dynamically by an ensemble of the units (see
Fig.1). Such dynamic circulation is an essential characteristic of a complex biological system\cite{CS,CSB}: genes encoded in the DNA control macroscopic phenotypes in an organism, while competition between phenotypes at the population level determines the expression of genes.

This interdependence between hierarchy levels has been studied in statistical physics, in particular, in collective phenomena. Self-consistent solutions or approximations are usually adopted in studying cooperative phenomena, where stationary, consistent relationships between microscopic elements and their mean (collective) field are generated. Although statistical physics is important for studying complex biological systems, an essential factor in biological systems is not addressed in standard statistical physics: A biological unit usually has the potential to reproduce. With reproduction, the number of units increases, which may change the relationship between levels, since the upper level consists of a population of lower-level units, and this number changes in time. This may lead to instability in the consistency between elements and the mean (collective) field. Despite changes in population size, biological systems generally maintain a degree of consistency between levels, even though each unit has 
many degrees of internal freedoms (e.g., a cell has a huge variety of molecules). 

For example, in cell reproduction, the duplication of molecules in a cell is correlated such that it keeps some synchrony with the reproduction cycle of a cell. In the development of a multicellular organism, reproduction of cells is correlated so that the growth of each cell does not interfere with the growth of an ensemble of cells.

\begin{figure}
\centering
\includegraphics[height=4cm]{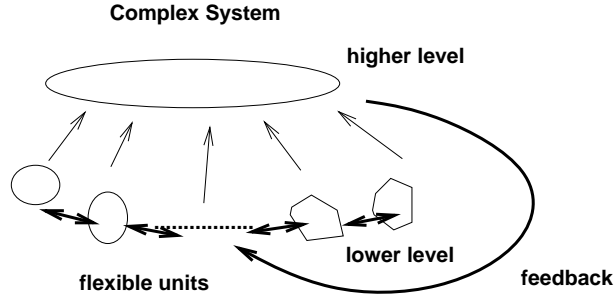}
\caption{Schematic picture for a hierarchical system.
In contrast to a simple system, a complex system is regulated by feedback from an upper to a lower level. Consistency between the levels should be considered.}
\end{figure}

Besides the potential for reproduction and internal reproduction, a biological unit often has the potential to evolve, which requires consistency between the time scales of evolution and of development of each unit. Phenotypes are generated as a result of the developmental process, which is robust both against noise in the developmental process and against some genetic mutations. Although the time scales of development and evolution are different, both are robust against noise or mutation, which suggests consistency between development and evolution\cite{Wagner1,Wagner2,Plausibility}.

Here we propose that understanding ``consistency'' among levels and its consequence is important for understanding a biological system. In question is how such consistency between different levels is sustained and whether there are resulting universal laws that apply to all biological systems.

Here we attempt to answer these questions by considering three examples: statistical laws
representing consistency between molecule replication and cell reproduction; general relationships between genetic variation and phenotypic fluctuation resulting from consistency between developmental and evolutionary stability; and the general robust cell differentiation process resulting from consistency between cell reproduction and multicellular development.

\section{Consistency between cell reproduction and molecule replication}

\subsection{Reaction network for cell reproduction}

A cell consists of several replicating molecular species that help in the synthesis of new molecules through catalytic reactions. As a result, a cell grows until it divides to produce two cells with similar chemical compositions (see Fig.2).

\begin{figure}
\centering
\includegraphics[height=4cm]{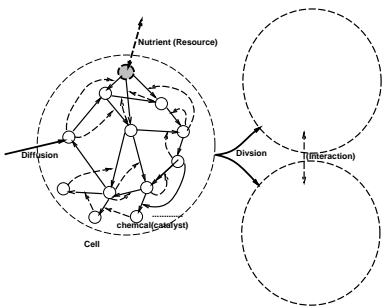}
\caption{Basic structure of a reproducing cell with internal catalytic chemical reactions. Cell--cell interaction is discussed in \S 4.}
\end{figure} 

These molecule replications must be synchronized; otherwise, the cellular chemical composition cannot be maintained, and the reproduction of cells with similar compositions will not continue. At the very least, a membrane to separate a cell from the environment must be synthesized, and this process must be synchronized with the replication of other internal
chemicals. It is unclear how such recursive production and chemical diversity can be maintained. Perhaps there is some statistical law for a system to sustain such reproduction.

To investigate the intracellular dynamics of replicating cells, we studied several cell models with intracellular catalytic reaction networks that transform nutrient chemicals into other chemical species\cite{Zipf} (see Fig.2 for a schematic representation). Within a cell there are a large number of chemical species that mutually catalyze a reaction. There are nutrient chemicals that are transported from the outside of a cell and are then transformed to other molecules through catalytic reactions. When the number of specific (or total) molecule species increases beyond a given threshold, the cell divides.

We studied a variety of models within this class, by adopting a
stochastic simulation for the reaction. By varying the speed of
nutrient intake, we found that the cell continues reproduction, 
keeping an approximately stable composition of chemicals, 
where the growth is optimized\cite{Zipf}. 
The recursive production is satisfied either by
tuning the diffusion constant $D$ of nutrient across the membrane, or
by introducing an active transport process \cite{SOC}. In the former
case, the fidelity of reproduction, i.e., the similarity of chemical
components is maximized at $D \rightarrow D_c$, where $D_c$ is a
critical point beyond which the production of a cell does not
continue. Around $D \sim D_c$, recursive production of a cell with
chemical diversity is possible.

\subsection{Universal power law in chemical abundances over species}
 
We investigated the universal statistical characteristics of reproduction state. First, we studied the statistics on the abundance of chemicals for a cell undergoing reproduction with constant chemical compositions. We measured the rank-ordered distributions of chemical species by plotting the number of molecules $n_i$ as a function of their rank as determined by $n_i$. As first reported in \cite{Zipf}, the distribution displays a power law with an exponent of --1, often called Zipf's law.

In our model, this power law of gene expression is maintained by a hierarchical organization of catalytic reactions. Major chemical species are synthesized through catalysis by less abundant chemicals. The latter chemicals are synthesized by chemicals with much less abundance, and this hierarchy of catalytic reactions continues until it reaches the most minor chemical species. Indeed, with the aid of mean-field analysis in statistical physics, we can show the appearance of a power law distribution with an exponent of --1.

We have confirmed the universality of Zipf's law by examining a variety of conditions in a cell model with reaction networks. They include (i) a distributed network connectivity such as the scale-free network distribution, (ii) distributed reaction coefficients, (iii) higher-order catalytic reactions, (iv) several schemes for transporting nutrients including active transport and passive diffusion, and (v) several schemes for cell divisions. Our results are constant despite modifications to the model. This power law in regard to abundances is observed in any cells that achieve recursive production, i.e., consistency between molecule replication and cell reproduction.

Furthermore, this power law is confirmed by measuring gene expressions (i.e., by measuring the abundances of a variety of mRNAs). For over a hundred cell types examined, we confirmed this power law with an exponent of --1 (see also \cite{Kuznetsov,Ueda}). The rank-ordered distribution of protein abundances in yeast cells, based on protein expression analysis shown in\cite{protein_abundance}, also suggests the power law with an exponent close to --1.

\subsection{Universal lognormal distribution of chemical abundances in cells}

We have thus far examined the average abundance of each chemical. Because the chemical reaction process is stochastic, the number of each type of molecule differs between cells. We therefore studied the distribution of each molecule number, sampled among cells, to find that the distribution is fitted reasonably well by the lognormal distribution, i.e.:

\begin{equation}
P(n_i) \sim \frac{1}{n_i} \exp(-\frac{(\log n_i-\overline{\log n_i})^2}{2\sigma}),
\end{equation}

\noindent
where $\overline{\log n_i}$ indicates the average of $\log n_i$ among cells. In other words, the distribution of the chemical abundances is fit by the normal (Gaussian) distribution, only after the logarithm of the abundances is taken. 

This lognormal distribution holds for the abundances of all chemicals, except for a few chemical species that are supplied externally to a cell as nutrients, which obey the standard Gaussian distribution\cite{Lognormal}. In other words, molecules that are reproduced in a cell obey the lognormal distributions. These results beg the question as to why the lognormal distribution law generally holds.

For illustration, consider an autocatalytic process where a molecule
(or a set of molecules) $x_m$ is replicated with the aid of other
molecules. Then, the growth of the number $n_m(t)$ of the molecule
species $x_m$ is given by
\begin{math}
dn_m(t)/dt=A n_m(t),
\end{math}
with $A$ describing the rates of the reaction processes that synthesize the molecule $x_m$.
Clearly, the synthetic reaction process depends on the number of the molecules involved in the catalytic process. At the same time, however, all chemical reaction processes are inevitably accompanied by fluctuations arising from the stochastic collisions of molecules.
Consequently, the above rate $A$ has fluctuations $\eta(t)$ around its temporal average $\overline{a}$ such that
\begin{math}
d n_m(t)/dt =n_m(t)(\overline{a} + \eta(t)),
\end{math}
and hence we obtain:
\begin{equation}
d\log n_m(t)/dt =\overline{a} + \eta(t).
\end{equation}

\noindent
In other words, the logarithm of the chemical abundances shows Brownian motion around its mean, as long as $\eta(t)$ is approximated by random noise. Accordingly, one would expect the logarithm of the chemical abundances (i.e., molecule numbers) to obey a normal (Gaussian) distribution.

In our cell model, however, each reaction process is not autocatalytic, but is catalytic with the aid of other molecule species, such that the equation is of the type $dn_i/dt = a n_j n_{\ell}$ ($i \neq \ell,j$). Thus, the above discussion cannot be directly applied. However, a multiplicative reaction process still exists. For example, it is the case that species 3 catalyzes the synthesis of species 1 and 2, as given by $dn_1/dt = a n_2 n_3$, and $dn_2/dt = a n_1 n_3$. Then, $d(n_1+n_2)/dt = a (n_1+n_2)n_3$ follows. If the molecule concentration $n_3$ fluctuates, the above argument on multiplicative noise is applied, leading to a lognormal distribution.

Of course, simple catalysis does not generally exist in our random catalytic network. We note a cascade reaction hierarchy, which supports the recursive production around the critical state $D \sim D_c$. A portion of possible reaction pathways are used dominantly, which organizes a cascade of catalytic reactions so that a chemical in the $i$-th group is catalyzed by the $(i+1)$-th, and that in the $(i+1)$-th group is catalyzed by the $(i+2)$-th, and so forth. A ``modular structure'' with groups of successive catalytic reactions is self-organized in the network. The fluctuations are successively multiplied through this cascade, i.e., the noise at the $(i+2)$-th level multiplicatively influences the $(i+1)$-th level, and $(i+1)$-th level to $i$-th level, $\cdots$, and so forth. By taking the logarithm of concentrations (i.e., $\log n_m$), these successive multiplications are transformed into successive additions of random variables, for which the central limit theorem is applied, leading to the Gaussian distribution of $\log n_m$.

Another simple derivation of lognormal-type distributions in the {\sl concentration}of chemicals is provided by noting that the change in concentration $c$ of some chemical in a cell is given by $dc/dt=$(Synthesis)-(Decomposition)-(Dilution), where the dilution results from the increase in cell volume $V$, given by the term $(dV/dt)c$. As the growth rate $dV/dt$ fluctuates, there appears a multiplicative noise term. By computing a stationary probability distribution of $c$ (from the Fokker--Planck equation corresponding to the Langevin equation), a log-tailed distribution appears (which approximately agrees with a lognormal distribution at a tail).

The lognormal distribution is observed in a variety of models where the cells reproduce efficiently. Following the above argument, the distribution reflects a balance between the replication of each molecule and the growth of the cell, leading to the equation $dn_m/dt$. This distribution is a result of consistency between replication of molecules and reproduction of a cell.

\subsection{Embedding the abundance power law into network topology}

Next, we investigated the relationship between the network connectivity statistics and the abundance statistics. The distributions in the connectivity of reaction networks has been studied extensively \cite{barabasi1,protein_network}, while the power law in chemical abundances discussed here is independent of the network structure, as long as the cell satisfies efficient and recursive growth.

The stability of reproduction as well as the growth speed may differ based on the network structure. We studied the evolution of the network by generating slightly modified networks and then selecting those that grew faster. We prepared $n$ cells with a randomly connected catalytic network with a given initial path number. Then, from each of these mother cells, $m$ mutant cells are generated by randomly adding one reaction path to the reaction network of the parent cell. Then, for each of the $n \times m$ cells, reaction dynamics are computed, to obtain the growth speed of each cell. Among the cell population, $n$ cells with higher growth speeds are selected. Again, from each of these cells, $m$ mutants are generated. This mutation--selection cycle is then repeated.

In the beginning, the parameters are not set at the critical point, such that Zipf's law on abundances is not observed. By using selection experiments to choose cells with higher growth speed, however, Zipf's law on abundances is achieved within 10 generations. The network structure is still random. When we continue selection processes, however, $P(k)$, the network connectivity distribution of reaction path numbers $k$, obeys a power law with an exponent close to --3\cite{PRE}.

This scale-free-type of connectivity distribution emerges in this evolution because attachment of paths to the chemicals with larger abundances is preferred as a result of the selection process. Note that the power law distribution of chemical abundances has already been established through evolution. A connection between a reaction path and a more abundant chemical is more effective in increasing the growth speed of a cell. A change in the growth speed by addition of an outgoing path from a given chemical species likely increases with its abundance $x$, since the degree of change increases in proportion to the flux of the reaction.
Then, the probability $q_{out}(x)$ to have such an outgoing path {\sl after selection} will increase with $x$, even though the addition of the path itself is random. If such probability linearly increases with $x$, then the abundance power law is transformed to the connectivity power law as $P(k_{out}) \propto {k_{out}}^{-2}$.
Numerically, we found that the probabilities $q_{out}(x)$ are fit by $q_{out}(x)\propto x^{\alpha}$ with $\alpha \approx 1/2$. In this case, the connectivity distribution is given by $k^{-(\alpha +1)/\alpha}=k^{-3}$\cite{PRE}.

It is interesting to note that the power law in abundances emerges first, and later, through evolution, it is embedded into the power law in the network connectivity. The abundance power law is in regard to the number of proteins, while the network is given by genes that ultimately determine whether a particular enzyme species that catalyzes a given reaction is present. When genes mutate, the network path is changed accordingly. In this sense, what we have observed here can be rephrased as ``phenotype (metabolic process or gene expression pattern) first, and genes later''. The power law in the abundance of the former is later ``assimilated'' by the gene network structure, as genetic assimilation as proposed by Waddington\cite{Waddington}.

\section{Consistency between genetic variation and phenotypic fluctuation}

\subsection{Evolutionary fluctuation response relationship}

The result of \S 2.3 suggests the existence of large phenotypic fluctuations among cells with identical genes. In the model, the network and the parameters are identical, and in the experiment, isogenic bacteria are used. Still, there exist large isogenic phenotypic fluctuations. Here we discuss the relevance of such fluctuations to evolution, in relation to genotype--phenotype mapping.

\begin{figure}
\noindent
\centering
\includegraphics[height=4cm]{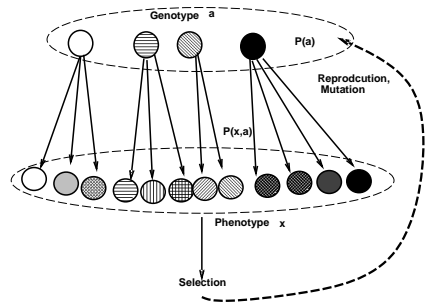}
\caption{Schematic representation of evolutionary processes with genotype and phenotype distributions.}
\end{figure}

Phenotypes are determined from genes through the developmental process. However, we note that the developmental process from a single genotype does not necessarily produce a single phenotype (see Fig.3). As mentioned, the developmental process is generally noisy, and, hence, the phenotype (such as abundances of proteins) from isogenic cells (organisms) fluctuates.

One might think that such isogenic phenotype fluctuations are not related to evolution, since phenotypic change without genetic change is not transferred to the next generation. However, the degree of fluctuation is determined by the gene, and is heritable. Hence, there may be a relationship between fluctuation and evolution. We have found evolutionary fluctuation--response relationships in bacterial evolution in the laboratory, where the evolution speed is proportional to the variance of the isogenic phenotypic fluctuation\cite{Sato}. This proportionality was confirmed in a simulation of the reaction network model in the last section\cite{JTB}. The origin of proportionality between the isogenic phenotypic fluctuation and genetic evolution has been discussed in light of the fluctuation--response relationship in statistical physics\cite{Einstein,Kubo}.

There is an established relationship between evolutionary speed and phenotypic fluctuation. It is the so-called fundamental theorem of natural selection by Fisher\cite{Fisher} that states that evolution speed is proportional to the variance of phenotype due to genetic variation, which is denoted as $V_g$. It is the phenotypic variance as a result of the distribution of genes in a population, defined as the fluctuation of variance of {\em average} phenotype over individuals with different genes. In contrast, the evolutionary fluctuation--response relationship, proposed here, concerns phenotypic fluctuation of isogenic individuals as denoted by $V_{ip}$. While $V_{ip}$ is defined as variance over clones, i.e., individuals with the same genes, $V_g$ is a result of the distribution of genes. Hence, the fluctuation--response relationship and the relationship concerning $V_g$ by Fisher's theorem are not identical.

If $V_{ip}$ and $V_g$ are proportional, the two relationships are consistent. Such proportionality, however, is not self-evident, as $V_{ip}$ is related to variation against the developmental noise and $V_g$ against the mutation. The relationship between the two, if it exists, postulates a constraint on genotype--phenotype mapping and may create a quantitative formulation of a relationship between development and evolution.

To determine this possible relationship, we again adopted the cell model with catalytic reaction networks and applied the genetic algorithm to evolve the network to increase a given fitness. Here, the fitness is given by the number $n_{i_s}$ of a given chemical $i_s$, so that reproducing cells with higher  $n_{i_s}$ are selected. We evolved cells (with recursive production as mentioned in \S 2), such that the concentration of a given chemical increases. We adopted a genetic algorithm with a fitness proportional to the concentration of $c_{i_s}$. Here, the mutation rate is given by the probability that a path in the network is added or deleted at each generation.

As mentioned, for a given network, there are fluctuations in the abundances of each chemical.
We took the phenotype variable $x=log(n_{i_s})$, since the distribution of $n_{i}$ is approximately lognormal, while theoretical studies \cite{Sato} adopt a variable $x$ whose distribution is close to Gaussian. As a measure of the phenotypic fluctuations, we computed a variance of $x$ for a network that gives peak abundances at each generation.

To investigate the distributions of phenotypes due to genetic variation, we computed the average phenotypes $\overline{x}$ over isogenic cells. This average phenotype $\overline{x}$ differs from mutant to mutant, from which its distribution is obtained. The variance of $\overline{x}$ over all mutants computed from this distribution gives $V_g$, while $V_{ip}$ is just the variance of $x$ from a given single genotype (i.e., network). 

\begin{figure} \noindent \centering
\includegraphics[height=5cm]{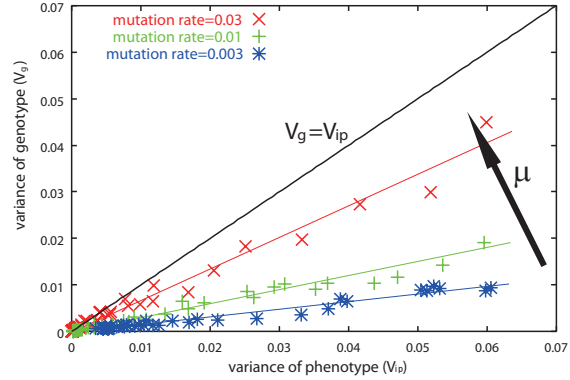}
\caption{The relationship between $V_{ip}$, the variance of the phenotype $x$ of the
isogenic cells and $V_g$, the variance of the phenotype $x$ over 10000 mutants of the selected cell. Plotted with identical symbols are the relationships over a given evolution course with fixed mutation rates as indicated in the figure (based on \cite{JTB}). } \end{figure}

$V_g$ versus $V_{ip}$ is plotted in Fig.4, which confirms that $V_{ip} \propto V_g$ holds for each evolutionary process with a fixed mutation rate. As the mutation rate $\mu$ increases, the slope of $V_g/V_{ip}$ increases, approaching the line $V_g =V_{ip}$. When $\mu$ is increased at some value $\mu_{max}$, mutant populations exhibiting very low values of $x$ increase, the distribution becomes flat, and the peak in the distribution shifts downwards. Indeed, around $\mu \approx \mu_{max}$, $V_g$ is the order of $V_{ip}$. 
For mutation rates beyond $\mu_{max}$, the phenotype distribution is almost flat, as shown in Fig.5, and the value of $x$ after selection cannot increase from generation to generation. Evolution no longer progresses. Thus, the evolution speed is optimal around $\mu \approx \mu_{max}$.

\begin{figure}
\noindent \centering
\includegraphics[height=4.5cm]{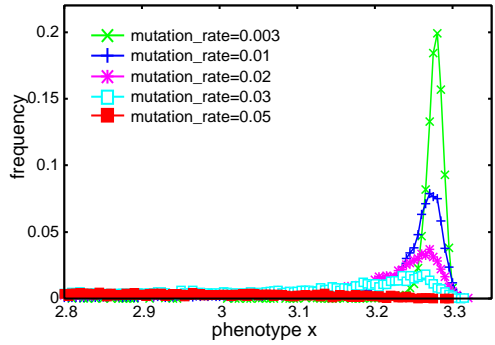}
\caption{Distribution of the phenotype $x$ over 10000 mutants, generated with mutation rates $0.003$, $0.01$, $0.02$, $0.03$, and $0.05$. When the mutation rate $\mu$ approaches $\mu_{max}$ (in this case, $\mu_{max}$ is around 0.03), the distribution is flattened, and the peak position shifts downward.}
\end{figure}

\subsection{Theoretical discussion}

Is it possible to formulate a phenomenological theory to support the relationship observed in numerical (and partially in in vivo) experiments presented in \S 3.1?

Here we consider the distribution both in phenotype $x$ and genotype $a$. Through the evolutionary process, the genotype changes from its dominant type $a=a_0$, and then the corresponding average phenotype for each genotype changes from $x=X_0$ accordingly. To investigate evolution, both with regards to the distribution of phenotype and genotype, we introduced a two-variable distribution, $P(x,a)$.

$V_{ip}$ is the variance of $x$, which can be written as $V_{ip}(a)=\int (x-\overline{x(a)})^2 P(x,a)dx$, where $\overline{x(a)}$ is the average phenotype of a clonal population sharing the genotype $a$, namely $\overline{x(a)}=\int P(x,a)x dx$.
$V_g$ is defined as the variance of the average $\overline{x(a)}$, over genetically heterogeneous individuals and is given by $V_{g}=\int (\overline{x(a)}-<\overline{x}>)^2 p(a)da$, where $p(a)$ is the distribution of genotype $a$ and $<\overline{x}>$ as the average of $\overline{x(a)}$ over all genotypes.

Assuming the Gaussian distribution, the distribution $P(x,a)$ is written as follows:

\begin{equation}
P(x,a)= \widehat{N}\exp[ -\frac{(x-X_0)^2}{2\alpha(a)}+C(a-a_0)(x-X_0)/\alpha-\frac{1}{2\mu}(a-a_0)^2],
\end{equation}

\noindent
With $\widehat{N}$ as a normalization constant. The Gaussian distribution $\exp(-\frac{1}{2\mu}(a-a_0)^2)$ represents the distribution of genotypes around $a=a_0$, whose variance is (in a suitable unit) the mutation rate $\mu$. The above equation can then be rewritten as:

\begin{equation}
P(x,a)= \widehat{N}\exp[-\frac{(x-X_0-C(a-a_0))^2}{2\alpha(a)}+(\frac{C^2}{2\alpha(a)}-\frac{1}{2\mu})(a-a_0)^2].
\end{equation}

Our second assumption {\bf evolutionary stability} states that at each stage of the evolutionary process, the distribution has a single peak in $(x,a)$ space. In order for this distribution to have a single peak (i.e., not to be flattened along the direction of $a$), the following condition (besides $\alpha >0$) should be satisfied:
\begin{math}
\frac{C^2}{2\alpha}-\frac{1}{2\mu} \leq 0,
\end{math}
i.e.,
\begin{math}
\mu \leq \frac{\alpha}{C^2} \equiv \mu_{max}.
\end{math}

\noindent
This means that the mutation rate has an upper bound $\mu_{max}$ beyond which the distribution does not have a peak in the genotype--phenotype space. Beyond this mutation rate, the distribution is extended to very low values of $x$ (fitness). This breakdown of the high-fitness phenotype is a kind of error catastrophe, if we follow the term by Eigen\cite{Eigen}. This is consistent with the observation in Fig.5.

We investigated the phenotypic variance due to the genotype distribution. First, we considered the average $\overline{x}_a$ over the distribution $P(x,a)$ for a given fixed $a$, and then considered the distribution of $\overline{x}_a$ according to the distribution $p(a)$,
noting that
\begin{math} 
\overline{x}_a \equiv \int x P(x,a) dx=X_0+C(a-a_0).
\end{math}
For the population having {\bf identical phenotype $x=X_0$}, the genetic variance in this distribution is given by $<(\delta a)^2> = \mu$. Hence, the phenotype variance from this population (of {\sl given phenotype} $x$) is written as:
\begin{equation} 
V_{ig}\equiv <(\overline{x}_a-\overline{x}_{a_0})^2> = C^2 \mu.
\end{equation}
We note that $C\neq 0$, so that the (average) phenotype changes with the change of genotype.
\noindent
Noting that $\alpha$ is the phenotypic variance $<\delta x^2>$ of isogenic individuals, $V_{ip}$, the inequality
\begin{math}
\mu<\alpha/C^2 \end{math}
is rewritten  as:
\begin{equation}
V_{ig} \leq V_{ip}.
\end{equation}

Since the genetic variance in the population $<(\delta a)^2>$ is proportional to the mutation rate, the above inequality is simply the threshold for the error catastrophe mentioned above. In other words, at the threshold mutation rate as $\mu_{max}$, $V_{ig} = V_{ip}$ holds. Then, recalling $V_{ig}\propto \mu$, we get
\begin{math}
V_{ig}=\frac{\mu}{\mu_{max}} V_{ip}.
\end{math}

In general, $V_g\neq V_{ig}$, as $V_g$ is the variance over {\sl all} current populations. In this case, the variance $<(\delta a)^2>$ is computed over the entire population, and is given by $\mu/(1-\mu C^2/\alpha)$. Thus, $V_g=V_{ig}/(1-V_{ig}/V_{ip})$. (Note $V_{ig}/V_{ip}<1$). If the mutation rate is small, however, $V_g \approx V_{ig}$, so that the proportionality between $V_{g}$ and $V_{ip}$ is explained from the above argument.

To sum up the present formulation of distribution,  we have obtained

(i) $V_{ip}\ \geq V_{ig}$, and (ii) proportionality between $V_g$ and $V_{ip}$ for small mutation rate cases through a given course of evolution are derived. 

Although these relationships are supported by the above simulation as well as by recent studies about gene networks\cite{Kaneko-robustness}, the above formulation is not a `derivation'.
First, we assumed the existence of two-variable distributions in genotype and phenotype $P(x,a)$. As genetic change is not given simply by the change of a continuous parameter, it is not a trivial assumption. (For example, the genetic change in the cell network model is an addition or a deletion of a reaction path, and if expressed as a continuous variable is not
self-evident). Second, the stability assumption assuring a single peak is expected to be valid for gradual evolution. Third, to adopt eq.(3), the existence of error catastrophe (to produce mutants with very low fitness values at a large mutation rate) is implicitly assumed.
With these three assumptions, error catastrophe is shown to occur at $V_{ig} \approx V_{ip}$.

In numerical evolution, the error catastrophe is estimated around $V_g \sim V_{ip}$. Here, the phenotype at each generation is within a small range, and the deviation of $V_{ig}$ from $V_g$ is not so large. Indeed, the estimate of the critical mutation rate for the error catastrophe is not accurate enough to distinguish between the two. Thus, the above theoretical estimate for the error catastrophe is consistent with the numerical result. \footnote{As $V_g \neq V_{ig}$, the inequality between $V_{ip}$ and $V_{ig}$ does not set the bound between $V_g$ and $V_{ip}$. Hence, the bound for heritability as anticipated at the discussion part of\cite{JTB} is not derived.}

The above results ask why the error catastrophe produces low-fitness phenotypes. Note that the growth of a cell in our model (and in nature) requires maintenance of a variety of chemicals through reproduction. Mutants may fail to synthesize some chemicals
concurrently. When the isogenic phenotypic fluctuation is large, there is room to search for networks that are robust against such mutational change, while for a large mutation rate, the network with the highest fitness is not maintained over generations.

The relationships (i)--(ii) as well as the estimate of error catastrophe are also confirmed in a gene network model\cite{Kaneko-robustness}. We expect these relationships (as well as the existence of the error catastrophe) to be generally valid for systems satisfying the following conditions.

(A) Fitness is determined through developmental dynamics.

(B) Developmental dynamics is complex such that its orbit, when deviated by noise, may fail to reach the state with the highest fitness.

(C) There is effective equivalence between mutation and noise in the developmental dynamics with regards to phenotype change.

Condition (A) is straightforward in our model, and condition (B) is satisfied because of the complex reaction dynamics to sustain the growth of a cell. Postulate (C) is satisfied since either the phenotypic fluctuation to increase the number of catalysts for a synthesis reaction or mutation to add another reaction term for its synthesis can contribute to the increase in the concentration of a given chemical in the same manner. Condition (A)-(B) supports the existence of `error catastrophe'. In several models of evolution, fitness is directly represented as a function of genotype. Then, a slight change in genotype does not lead to major differences in phenotype. On the other hand, in a system with (A)-(B), slight differences in genotype may lead to huge differences in phenotype, as the phenotype is determined after temporal integration of the developmental dynamics where slight differences in genotype (rule governing the dynamics) are accumulated and amplified. 

Note that relationship (ii) shows correlation between the phenotypic change by gene and spontaneous phenotypic fluctuation. In other words, the degree to which genes can alter phenotypes is presumed in the spontaneous phenotypic fluctuation. This suggests consistency between genetic and phenotypic levels (see Fig.3), as was first discussed by Waddington, as genetic assimilation\cite{Waddington}.

\section{Consistency between cell replication and reproduction of multicellular organisms}

We briefly discuss cell differentiation, i.e., diversification into a discrete set of cell types through development and robustness in the population distribution of each cell type through development.

\begin{figure}
\noindent \centering
\includegraphics[height=4.5cm]{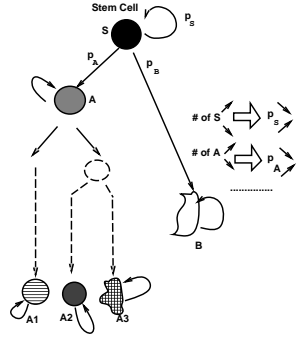}
\caption{Schematic representation of hierarchical differentiation from a stem cell.}
\end{figure} 

In cell differentiation in multicellular organisms, there usually exists irreversibility. Embryonic stem cells have the potential to produce all types of cells. Stem cells produced from them have the potential to produce only a limited class of cells (for example, a blood stem cell can only generate all types of blood cells). The stem cells proliferate or differentiate into other types with some probability. Successive differentiation from stem cells ($S$) terminates at determined cells, which can only replicate (see Fig.6). This cell society must then maintain some ratio of population sizes of cell types. Stem cells must then regulate the ratio of proliferation and differentiation, depending on the population distribution of each cell type. In the example displayed in Fig.6, the ratio of differentiation $S\rightarrow A$ has to be increased when the number of type A cells is decreased, while the ratio of proliferation $S\rightarrow S$ is increased when the number $S$ is decreased.

We have studied a reaction network model as given in Fig.2, choosing the model so that the chemical concentration changes chaotically\cite{stem}. We considered cell--cell interactions. As the cell number increases by cell divisions, the cells share common resources. Through the concentration change of resources in the medium, cells interact with each other. This interaction introduces a change in intracellular dynamics. 

Here, cell state is given by the composition of chemicals (i.e., by location in the phase space of chemical concentrations). Each cell type is determined as an `attractor' in the intracellular reaction dynamics during cell--cell interaction. Initially, the cellular state is a chaotic attractor that gives type `S'. With cell divisions, we found that some cells differentiate from the original cell state `S' to a different type with a distinct chemical composition (`A'), while another type (`B') is differentiated later. With further increases in the cell number, differentiation from A to A1, A2, and A3 progresses. In summary, cell differentiation progresses as displayed in Fig.6. Stem cells differentiate into other types stochastically due to the instability caused by cell--cell interactions.

The differentiation is `stochastic', arising from chaotic intracellular chemical dynamics. The choice of a stem cell either to proliferate or to differentiate appears stochastic as far as the cell type is concerned. However, this is not due to external fluctuations but is a result of the intracellular state. As this state is influenced by cell--cell interaction, the probability of differentiation can be regulated according to the population of cell types. As the number (fraction) of stem cells increases, the basin for such chaotic attractors touches with the attractor itself so that the cells switch to a different cellular state. On the other hand, as the fraction of stem cells decreases, the original chaotic attractor is stable so that the stem cells can proliferate.
With this spontaneous tuning in the stability of the stem cell state, the fraction of each cell type is regulated. On the other hand, the state of stem cells is `marginally stable', in the sense that its attractor is on the verge of touching with its basin, through this regulation.

Reaction networks allowing for chaotic dynamics and differentiation are not common. We checked a variety of networks to examine the growth speed of each cell (i.e., the inverse of the division time) and the ensemble of cells\cite{PRL}. First, cells having networks allowing for
chaotic dynamics and differentiation do not grow quickly at the cellular level. Cells having other networks without any oscillatory dynamics in chemical concentrations often divide faster. However, as the number of cells grows, the speed of division for such non-oscillating cells is drastically reduced, while for cells with chaotic dynamics and differentiation, the speed is not so reduced. This is because cells of the latter type do not strongly compete for resources. In other words, consistency between a single cell division and the growth of an ensemble of cells is achieved for cells having networks allowing for chaotic dynamics. Hence, the consistency between reproduction of each cell type and population growth as an ensemble of cells works as a pressure to select networks exhibiting chaotic dynamics. Regulation to maintain the proportion of each cell type and the `marginal stability' of the stem cell state are a consequence of the cell system that satisfies the consistency.

\section{Conclusion}

Here we reviewed three problems in biology from the viewpoint of `consistency between different levels'. First, as a result of consistency between molecule replication and cell reproduction, chemical reaction dynamics are shown to be at a critical state, and a power law distribution of chemical abundances (gene expression) is derived. The dynamics of the molecule number is shown to include a multiplicative noise term, which leads to a lognormal distribution of chemical abundances over cells. Through evolution, the power law distribution of abundances is embedded into the network topology, leading to a scale-free network, demonstrating consistency between reaction dynamics and network structure. 

Second, the genotype--phenotype relationship is discussed. We found a general relationship between phenotype fluctuations by genetic variation, $V_g$, and isogenic phenotypic fluctuation by developmental noise, $V_{ip}$, as a result of a consistency between genotype and phenotype, because of feedback from phenotype to gene through selection. 

As the third topic, we touched upon the chaos mechanism for stem cell differentiation with autonomous regulation, as a result of consistency between cell reproduction and growth of the cell ensemble. Although not described here, recent studies show the existence of spontaneous adaptation as a result of consistency between cellular growth and (stochastic) gene expression patterns\cite{Kashiwagi,PLOSCOMP}. We propose that the consistency principle is generally relevant to understanding reproduction, adaptation, evolution, and development in biological systems.

The authors would like to S. Sawai, M. Tachikawa, K. Fujimoto, and T. Yomo for stimulating discussions.

\end{document}